\newtheorem{theorem}{Theorem}
\newtheorem{example}{Example}
\newtheorem{corollary}{Corollary}
\newtheorem{lemma}{Lemma}
\newtheorem{definition}{Definition}

\newtheorem{proposition}{Proposition}

\def\psfancypar#1#2{\begingroup\def\par{\endgraf\endgroup\lineskiplimit=0pt}
               \setbox2=\hbox{\large\sc #2}
               \newdimen\tmpht \tmpht \ht2 \advance\tmpht by \baselineskip
               \font\hhuge=Times-Bold at \tmpht
               \setbox1=\hbox{{\hhuge #1}}
               \count7=\tmpht \count8=\ht1
               \divide\count8 by 1000 \divide\count7 by \count8
               \tmpht=.001\tmpht\multiply\tmpht by \count7
               \font\hhuge=Times-Bold at \tmpht
               \setbox1=\hbox{{\hhuge #1}}
               \noindent
                \hangindent1.05\wd1
               \hangafter=-2 {\hskip-\hangindent
               \lower1\ht1\hbox{\raise1.0\ht2\copy1}%
                \kern-0\wd1}\copy2\lineskiplimit=-1000pt}

\newcommand{\beq}{\begin{equation}}
\newcommand{\eeq}{\end{equation}}
\newcommand{\bqa}{\begin{eqnarray}}
\newcommand{\eqa}{\end{eqnarray}}
\newcommand{\bqn}{\begin{eqnarray*}}
\newcommand{\eqn}{\end{eqnarray*}}
\newcommand{\nn}{\nonumber}

\newcommand{\be}{\begin{enumerate}}
\newcommand{\ee}{\end{enumerate}}
\newcommand{\bi}{\begin{itemize}}
\newcommand{\ei}{\end{itemize}}
\newcommand{\bd}{\begin{description}}
\newcommand{\ed}{\end{description}}
\newcommand{\ba}{\begin{array}}
\newcommand{\ea}{\end{array}}
\newcommand{\bde}{\begin{definition}}
\newcommand{\ede}{\end{definition}}
\newcommand{\bex}{\begin{example}}
\newcommand{\eex}{\end{example}}

\def\boxit#1{\vbox{\hrule\hbox{\vrule\kern3pt
        \vbox{\kern3pt#1\kern3pt}\kern3pt\vrule}\hrule}}

\def\reals{ { {\rm  I \kern-0.15em R }  } }
\def\complex{ {\,{{\rm C} \kern-0.50em \raise0.20ex {  |}}\, }}

\def\0bf{{\bf 0}}
\def\1bf{{\bf 1}}
\def\2bf{{\bf 2}}
\def\3bf{{\bf 3}}
\def\4bf{{\bf 4}}
\def\5bf{{\bf 5}}
\def\6bf{{\bf 6}}
\def\7bf{{\bf 7}}
\def\8bf{{\bf 8}}
\def\9bf{{\bf 9}}

\def\Rbf{{\bf R}}

\def\Tmat{\mathcal{T}}

\def\Rxx{\Rbf_{\ssstyle X\kern-.1em X}}

\let\ssstyle=\scriptscriptstyle

\def\Kout{\setbox1=\hbox{\Huge\bf K}\hbox to
1.05\wd1{\hspace{.05\wd1}
}}
\def\Sout{\setbox1=\hbox{\Huge\bf S}\hbox to 1.05\wd1{\hspace{.05\wd1}
}}

\documentclass[12pt,draft,onecolumn]{IEEEtran}
\usepackage{epsf,psfrag,amssymb,amsfonts,cite}
\usepackage{graphicx}
\def\scalefig#1{\epsfxsize #1\textwidth}
\begin{document}
\title{\Large \bf Interference Channels With Arbitrarily Correlated Sources }
\IEEEoverridecommandlockouts
\author{\authorblockN{Wei Liu and Biao Chen}\\
\authorblockA{Department of EECS, Syracuse University\\
Syracuse, NY 13244\\
Email: wliu28\{bichen\}@syr.edu}
\thanks{This work was supported by the National Science Foundation under Grants CCF-0546491 and CCF-0905320. The material in
this paper was presented in part at the 47th Annual Allerton Conference on Communication, Control, and Computing, Montecillo, IL, October 2009.}}

\maketitle
\begin{abstract}
Communicating arbitrarily correlated sources over interference channels is considered in this paper. A sufficient condition is found for the lossless transmission of a pair of correlated sources over a discrete memoryless interference channel. With independent sources, the sufficient condition reduces to the Han-Kobayashi achievable rate region for the interference channel. For a special correlation structure (in the sense of Slepian-Wolf, 1973), the proposed region reduces to the known achievable region for interference channels with common information. A simple example is given to show that the separation approach, with Slepian-Wolf encoding followed by optimal channel coding, is strictly suboptimal.
\end{abstract}
\section{Introduction}
Communicating correlated sources over multi-terminal networks have been a topic of research interest in the past decades. Slepian and Wolf \cite{Slepian&Wolf:73IT2} studied communicating correlated information over a two-user multiple access channel where the correlation is of a special structure in the form of three independent sources, with one of them observed by both encoders while each of the other two observed only at individual encoders. Later, Cover, El-Gamal, and Salehi studied the problem of communicating discrete correlated sources over a multiple access channel (MAC) \cite{Cover_etal:1980}, where the correlation structure can be arbitrary. A sufficient condition was obtained for the lossless transmission of such correlated source pair that includes various known capacity results as its special cases. These include the capacity region for a MAC \cite{Ahlswede:71ISIT,Liao:thesis}; distributed lossless source coding, i.e., the Slepian-Wolf coding\cite{Slepian&Wolf:73IT}; cooperative multiple access channel capacity; and the correlated source multiple access channel capacity region of Slepian and Wolf\cite{Slepian&Wolf:73IT2}. The key technique used in \cite{Cover_etal:1980}, aside from making use of the so-called common part of correlated random variables (in the sense of G\'{a}cs, K\"{o}rner \cite{Gacs&Korner:1973} and Witsenhausen\cite{Witsenhausen:1975}), is the correlation preserving codeword generation. By generating codewords that depend, probabilistically, on the source sequences, the correlation between the sources induces correlation in the generated codewords. In addition, a simple example was given to show that the separation approach which concatenates a Slepian-Wolf code \cite{Slepian&Wolf:73IT} and the optimal channel code for MAC \cite{Ahlswede:71ISIT,Liao:thesis} is strictly suboptimal.

Han and Costa \cite{Han&Costa:1987} studied the problem of communicating  arbitrarily correlated sources over a discrete memoryless broadcast channel. The sufficient condition derived in \cite{Han&Costa:1987} (with correction by Kramer and Nair \cite{Kramer&Nair:09ISIT}) recovers the Marton region for broadcast channels with independent messages \cite[Theorem 2]{Marton:1979}. In \cite{Minero&Kim:09ISIT}, Minero and Kim proposed an alternative coding scheme and the obtained region was shown to be equivalent to that of Han and Costa. In addition, it was pointed out in \cite{Minero&Kim:09ISIT} that the common part does not play a role for the broadcast channel case which is consistent with the engineering intuition because of the centralized transmitter. We comment here that the same coding scheme proposed by Han and Costa can also be easily modified to obtain the same region without the use of the common part, as to be elaborated in Section II.

Communicating correlated sources over interference channels has previously been studied by Salehi and Kurtas \cite{Salehi&Kurtas:93ISIT}. However, the obtained rate region, derived by largely following the coding scheme for the MAC channel \cite{Cover_etal:1980,Ahlswede&Han:83IT} does not reduce to the well known Han and Kobayashi (HK) region for interference channels \cite{Han&Kobayashi:81IT} when the sources are independent. We remark here that the HK region, originally proposed in 1981 \cite{Han&Kobayashi:81IT} and recently simplified by Chong {\em et al}\cite{Chong_etal:2008}, remains to be the largest achievable rate region for interference channels with independent messages. In addition, there is no definitive answer to the question whether the separation approach is strictly suboptimal, even though intuition suggests that this is likely the case.

In this work, we derive a sufficient condition for the lossless transmission of a pair of arbitrarily correlated sources over a discrete memoryless interference channel (DMIC). The coding scheme takes advantage the common part of the random source pair, if it exists. Moreover, it utilizes the correlation preserving technique for the multiple access channel \cite{Cover_etal:1980} and the random source partition for the broadcast channel \cite{Han&Costa:1987}. We show that the proposed region includes the HK region as its special case. In addition, for a special correlation structure (in the sense of Slepian-Wolf, 1973 \cite{Slepian&Wolf:73IT2}), the proposed region coincides with the known achievable region for interference channels with common information \cite{Jiang-etal:06Arsilomar,Cao&Chen&Zhang:WCNC07,Jiang&Xin:2008}. We also give a simple example to show that the separation approach for communicating correlated sources over interference channels is strictly suboptimal.

The rest of this paper is organized as follows. Section II gives the problem formulation and introduces some previous results related to this work.
The main results are presented in Section III. In Section IV, a simple example is given to show that separation is strictly suboptimal for communicating correlated sources over interference channels.  Section V  concludes this paper.
\section{Definitions, Problem statement and Related works}
\begin{figure}
\centerline{
\begin{psfrags}
\psfrag{sn}[c]{$s^n$}
\psfrag{tn}[c]{$t^n$}
\psfrag{en1}[c]{\quad \ \ Encoder 1}
\psfrag{en2}[c]{\quad Encoder 2}
\psfrag{x1n}[c]{$x^n_1$}
\psfrag{x2n}[c]{$x^n_2$}
\psfrag{py1y2}[l]{$p(y_1y_2|x_1x_2)$}
\psfrag{y1n}[c]{$y^n_1$}
\psfrag{y2n}[c]{$y^n_2$}
\psfrag{de1}[c]{\quad \ \ Decoder 1}
\psfrag{de2}[c]{\quad \ \ Decoder 2}
\psfrag{shatn}{$\hat{s}^n$}
\psfrag{thatn}{$\hat{t}^n$} \scalefig{.8}\epsfbox{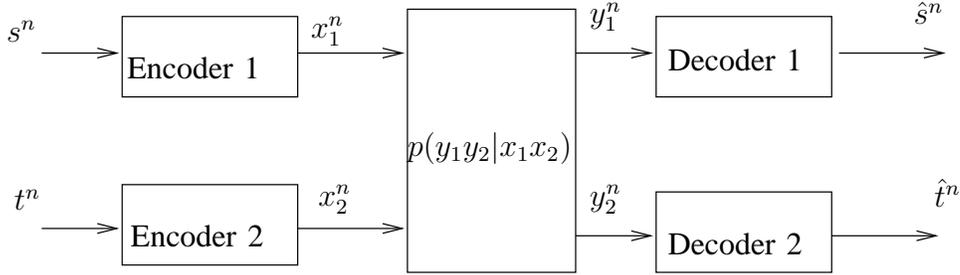}
\end{psfrags}}
\caption{\label{fig:model} Interference channels with correlated sources.}
\end{figure}

The model studied in this paper is shown in Fig. \ref{fig:model}. The source sequences $(S^n, T^n)$ are arbitrarily correlated discrete memoryless sources, generated independently according to:
\beq p(s^n,t^n)=\prod^n_{i=1}p(s_i,t_i).\eeq 
This pair of source sequences $S^n$ and $T^n$ are to be transmitted losslessly over a two user discrete memoryless interference channel
defined by the transition probability $p(y_1y_2|x_1x_2)$, where $X_1, X_2$ are the channel inputs and $Y_1, Y_2$ are the channel outputs.

A length $n$ source channel block code for the channel consists of two encoder mappings:
\bqa f_1&:& {\cal S}^n\rightarrow {\cal X}^n_1,\\
     f_2&:& {\cal T}^n\rightarrow {\cal X}^n_2,\eqa
 and two decoder mappings:
 \bqa \phi_1&:& {\cal Y}^n_1\rightarrow {\cal S}^n,\\
    \phi_2&:& {\cal Y}^n_2\rightarrow {\cal T}^n.\eqa

 The probability of error at decoders 1 and 2 are defined as
 \bqa P_{e1}&=&\sum_{s^n\in{\cal S}^n}p(s^n)Pr\{s^n\neq d_1(y^n_1)|S^n=s^n\},\\
 P_{e2}&=&\sum_{t^n\in{\cal T}^n}p(t^n)Pr\{t^n\neq d_2(y^n_2)|T^n=t^n\}.\eqa


\begin{definition}
The source $(S, T)\sim \prod^n_{i=1}p(s_i,t_i)$ is said to be {\em admissible} for the interference channel $p(y_1y_2|x_1x_2)$ if for any $\epsilon\in(0,1)$ and sufficiently large $n$, there exist a sequence of block codes $(f_1,f_2,\phi_1,\phi_2)$ such that
\bqa \max\{P_{e1},P_{e2}\}\leq \epsilon. \eqa
\end{definition}

The goal of this paper is to find a sufficient condition  such a source $(S, T)$ is admissible for a given DMIC. In the following, we summarize some previous results related to this work.

Cover, El Gamal, and Salehi \cite{Cover_etal:1980} obtained the following sufficient condition for the lossless transmission of arbitrarily correlated sources over a multiple access channel.
 \begin{proposition}(\cite[Theorem 1]{Cover_etal:1980})  A source pair $(S^n,T^n)\sim \prod^n_{i=1}p(s_i,t_i)$ can be sent with arbitrarily small probability of error over a discrete memoryless multiple access channel $p(y|x_1,x_2)$ if
    \bqa H(S|T)&<& I(X_1;Y|X_2SW),\\
     H(T|S)&<& I(X_2;Y|X_1TW),\\
     H(ST|K)&<&I(X_1X_2;Y|KW),\\
     H(ST)&<&I(X_1X_2;Y),\eqa
     where \beq p(s,t,w,x_1,x_2,y)=p(w)p(s,t)p(x_1|sw)p(x_2|tw)p(y|x_1x_2), \eeq
     and $K=f(S)=g(T)$ is the common part of two variables $(S,T)$, in the sense of G\'{a}cs, K\"{o}rner \cite{Gacs&Korner:1973} and Witsenhausen\cite{Witsenhausen:1975}.
     \end{proposition}
     
    The key technique in deriving the sufficient condition is the correlation preserving codeword generation. For fixed distribution $p(w), p(x_1|w,s), p(x_2|w,t)$, independently generate one codeword $w^n(k^n)\sim \prod^n_{i=1}p(w_i)$ for each $k^n\in{\cal K}^n$, to carry the information of the common part. Next, for each source sequence $s^n\in {\cal S}^n$, find the corresponding $k^n=f(s^n)=(f(s_1),f(s_2),\cdots, f(s_n))$ and independently generate one codeword $x^n_1\sim \prod^n_{i=1}p(x_{1i}|s_i,w_i)$. The codeword $x_2^n$ is similarly generated. Therefore the correlation between the sources induces correlation in the generated codewords, the so-called correlation preserving codeword generation. To transmit $s^n$, encoder 1 sends the corresponding codeword $x^n_1$. Similarly  encoder 2  sends the corresponding codeword $x_2^n$ for the given source sequence $t^n\in \Tmat^n$. The decoder uses joint typicality decoding: upon receiving $y^n$, the decoder finds a unique pair of $(s^n,t^n)$ such that $(s^n,t^n,k^n,w^n,x^n_1,x^n_2,y^n)\in T^n_{\epsilon}(STKWX_1X_2Y)$.

Han and Costa \cite{Han&Costa:1987} proposed the following sufficient condition for the lossless transmission of arbitrarily correlated sources over a discrete memoryless broadcast channel.

\begin{proposition}(Han and Costa\cite{Han&Costa:1987}, with correction by Kramer and Nair\cite{Kramer&Nair:09ISIT})  A source pair $(S^n,T^n)\sim \prod^n_{i=1}p(s_i,t_i)$ can be sent with arbitrarily small probability of error over a discrete memoryless broadcast channel $p(y_{1}y_{2}|x)$ if there exist auxiliary random variables $W, U, V$ satisfying the Markov chain property $ST \rightarrow WUV \rightarrow X \rightarrow Y_1Y_2$ such that
    \bqa H(S)&<& I(SWU;Y_1)-I(T;WU|S),\\
     H(T)&<& I(TWV;Y_2)-I(S;WV|T),\\
     H(ST)&<&\min\{I(KW;Y_1),I(KW;Y_2)\}+I(SU;Y_1|KW),\\
             &&+I(TV;Y_2|KW)-I(SU;TV|KW),\\
     H(ST)&<&I(SWU;Y_1)+I(TWV;Y_2)-I(SU;TV|KW)-I(ST;KW),\eqa
           where $K=f(S)=g(T)$ is the common part of the two variables.
\end{proposition}

The key technique that is of particular use to our problem is the random source partition, which reminisces superposition coding for the channel coding problem.
Specifically, source sequences $s^n \in {\cal S}^n$, $t^n\in {\cal T}^n$ are randomly placed into $2^{nr_1}$ and $2^{nr_2}$ cells, respectively. The cell indices for $s^n$ and $t_n$, denoted by $\alpha$ and $\beta$, respectively,  play the role as  the common information to be decoded by both receivers. The coding scheme is sketched as follows: fix distribution $p(w), p(u|w,s)$, and $p(v|w,t)$. For each $\alpha, \beta$ and $k^n$, independently generate $2^{n\rho_0}$ codewords $w^n(\alpha,\beta,k^n)\sim \prod^n_{i=1}p(w_i)$. Next, for each pair of $(s^n, w^n)$, independently generate $2^{n\rho_1}$ codewords $u^n(s^n,w^n)\sim \prod^n_{i=1}p(u_i|s_i,w_i)$, and $2^{n\rho_2}$ codewords $v^n(t^n,w^n)\sim \prod^n_{i=1}p(v_i|t_i,w_i)$. For each pair of source sequences $(s^n,t^n)$, the encoder will choose a triple $(w^n, u^n, v^n)$ such that $(s^n, t^n, k^n, w^n,u^n,v^n)\in T^n_{\epsilon}(STKWUV)$, which is ensured with high probability by properly chosen $\rho_1,\rho_2$ and $\rho_3$. The two decoders use joint typicality decoding, that is, decoder $Y_1$ will find a unique sequence $s^n$ such that $(s^n,k^n,w^n,u^n,y^n_1)\in T^n_{\epsilon}(SKWUY_1)$. Similarly, decoder $Y_2$
will find a unique sequence $t^n$ such that $(t^n,k^n,w^n,v^n,y^n_2)\in T^n_{\epsilon}(TKWVY_2)$.

In \cite{Minero&Kim:09ISIT}, Minero and Kim proposed an alternative, and conceptually simple, coding scheme. The obtained region does not involve the common part $K$ of the two variables $(S, T)$, but was shown to be equivalent to that of Han and Costa. In addition, it was pointed out in \cite{Minero&Kim:09ISIT} that the common part does not play a role for the broadcast channel case which is consistent with the engineering intuition because of the centralized transmitter. Indeed, the same coding scheme proposed by Han and Costa can also be easily modified to obtain the same region without the use of the common part. For the encoding scheme in \cite{Han&Costa:1987}, sketched above, remove the part related to the common variable $K$, in both the encoding and decoding processes, straightforward error probability analysis leads to the following sufficient conditions:
\bqa H(S)&<& I(SWU;Y_1)-I(T;WU|S),\\
     H(T)&<& I(TWV;Y_2)-I(S;WV|T),\\
     H(ST)&<&\min\{I(W;Y_1),I(W;Y_2)\}+I(SU;Y_1|W),\\
             &&+I(TV;Y_2|W)-I(SU;TV|W),\\
     H(ST)&<&I(SWU;Y_1)+I(TWV;Y_2)-I(SU;TV|W)\nn\\
     &&-I(ST;W),\eqa
     where $W, U, V$ satisfying the Markov chain property $ST \rightarrow WUV \rightarrow X \rightarrow Y_1Y_2$.
     
This region is the same as in \cite[Theorem 3.1]{Minero&Kim:09ISIT}, which was shown to be equivalent to that described in Proposition 2.

For interference channels with independent messages, the largest achievable rate region was given by Han and Kobayashi \cite{Han&Kobayashi:81IT}. The HK region was recently simplified by Chong {\em et al} and we repeat in Proposition 3 this simplified HK inner bound.

\begin{proposition}(Chong {\em et al}\cite[Theorem 2]{Chong_etal:2008}) Let ${\cal P}$ be the set of probability distributions that factor as
$ P(q,w_1,w_2,x_1,x_2)=p(q)p(w_1|q)p(w_2|q)p(x_1|w_1q)p(x_2|w_2q)$.
Then the rate pair $(R_1,R_2)$ is achievable for a discrete memoryless interference channel $p(y_1y_2|x_1x_2)$, if the following conditions are satisfied:
 \bqa R_1&<&I(X_1;Y_1|QW_2),\\
R_2&<&I(X_2;Y_2|QW_1),\\
R_1+R_2&<&I(X_1;Y_1|QW_1W_2)+I(W_1X_2;Y_2|Q),\\
R_1+R_2&<&I(X_2;Y_2|QW_1W_2)+I(W_2X_1;Y_1|Q),\\
R_1+R_2&<&I(W_2X_1;Y_1|QW_1)+I(W_1X_2;Y_2|QW_2),\\
2R_1+R_2&<&I(X_1;Y_1|QW_1W_2)+\nn\\
&&I(W_2X_1;Y_1|Q)+I(W_1X_2;Y_2|QW_2),\\
R_1+2R_2&<&I(X_2;Y_2|QW_1W_2)+\nn\\
&&I(W_1X_2;Y_2|Q)+I(W_2X_1;Y_1|QW_1).\eqa
\end{proposition}

\section{Main results}

 We start with a quick review of the HK achievable rate region of interference channel with independent messages. The major ingredients in the coding scheme for the HK region is rate splitting and joint decoding. Specifically, user $i$, $i=1,2$, splits the message $M_i$ into two parts, common message $M_{i0}$ and private message $M_{i1}$. Therefore, $|M_i|=|M_{i0}|\times |M_{i1}|$ where $|\cdot|$ denotes the cardinality of a set. The common message needs to be decoded by both decoders and the private message is only intended for its own receiver. This rate splitting can be implemented using sequential superposition encoding as described in \cite{Chong_etal:2008}. Let $R_{ij}=\frac{1}{n}\log |M_{ij}|$, $i=1,2$ and $j=0,1$. First generate $2^{nR_{i0}}$ auxiliary codewords $w^n_i$, which carry the information of common message $M_{i0}$. Next, for each $W^n_{i}$, generate $2^{nR_{i1}}$ codewords $x^n_i$ superimposed on top of $w^n_i$, which carry the information of the private message $M_{i1}$. Each decoder jointly decodes both common messages and its own private message, i.e., decoder 1 finds unique codewords $w^n_1$, $w^n_2$ and $x^n_1$ such that $(w^n_1,w^n_2,x^n_1,y^n_1)\in T^n_{\epsilon}(W_1W_2X_1Y_1)$, and  decoder 2 finds unique codewords $w^n_1, w^n_2$ and $ x^n_2$ such that $(w^n_1,w^n_2,x^n_2,y^n_2)\in T^n_{\epsilon}(W_1W_2X_2Y_2)$.

\begin{figure}
\centerline{\begin{psfrags}
\psfrag{sn}[c]{$s^{n}$}
\psfrag{tn}[c]{$t^n$}
\psfrag{kn}[c]{$k^n=f(s^n)=g(t^n)$}
\psfrag{wn0}[c]{$w^n_0(k^n)$}
\psfrag{v}[c]{$\vdots$}
\psfrag{cdot}[c]{$\cdots$}
\psfrag{a}[c]{$\alpha$}
\psfrag{b}[c]{$\beta$}
\psfrag{wn1}[c]{$w^n_1(\alpha,w^n_0(k^n))$}
\psfrag{wn2}[c]{$w^n_2(\beta,w^n_0(k^n))$}
\psfrag{pw1}[c]{$p(w_1|w_0)$}
\psfrag{pw2}[c]{$p(w_2|w_0)$}
\psfrag{px1}[c]{$p(x_1|sw_1w_0)$}
\psfrag{px2}[c]{$p(x_2|tw_2w_0)$}
\psfrag{1}[c]{$1$}
\psfrag{r1}[c]{$2^{nr_1}$}
\psfrag{r2}[c]{$2^{nr_2}$}
\psfrag{x1}[c]{$x^n_1(s^n,w^n_1,w^n_0)$}
\psfrag{x2}[c]{$x^n_2(t^n,w^n_2,w^n_0)$}
\scalefig{0.9}\epsfbox{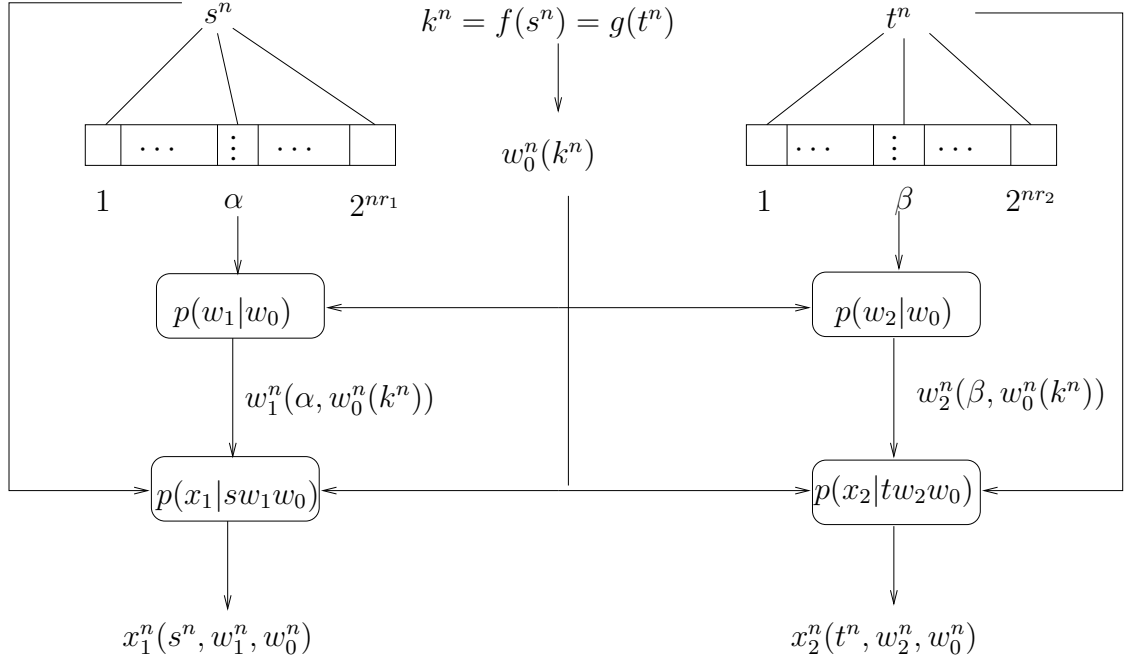}
\end{psfrags}}
\caption{\label{fig:ic1} Coding structure for IC with correlated sources}
\end{figure}

Consider now the model of interest in the present paper, i.e., IC with correlated sources. Let us first disregard the common part $K$ between the source variables $S$ and $T$. We start with Han and Costa's random source partition: the sequences  $s^n\in {\cal S}^n$ and $t^n \in {\cal T}^n$ are randomly placed respectively into $2^{nr_1}$ and $2^{nr_2}$ cells.  This source partition is tantamount to rate splitting in the channel coding problem: the cell index associated with a given sequence plays the role of common information and the index of the source within the cell the private information. This is then followed by the superposition coding \cite{Chong_etal:2008}. First generate an auxiliary codeword $w_i^n$ for each cell index. The codeword $x_i^n$ is then generated to be superimposed on top of $w_i^n$ that also carries the source index within the cell. Different from \cite{Chong_etal:2008} is that the codeword $x_i^n$ is statistically dependent on the input source, thereby preserving the correlation contained in the original source pair. The common part $K$, if it exists, can then be put back in the encoding process by generating an auxiliary codeword $w_0^n$. This codeword, known to both encoders, will be used in generating all the other codewords through a superposition code structure. This encoding process is illustrated in Fig. \ref{fig:ic1}.

For decoders, joint typicality decoding is used at both decoders. That is, decoder 1 finds a unique $s^n$ such that $(s^n,k^n,w^n_0,w^n_1,x^n_1,w^n_2,y^n_1)\in T^{\epsilon}_n(SKW_0W_1X_1W_2Y_1)$, and  decoder 2 finds a unique $t^n$ such that $(t^n,k^n,w^n_0,w^n_2,x^n_2,w^n_1,y^n_2)\in
T^{\epsilon}_n(TKW_0W_2X_2W_1Y_2)$. The above coding scheme leads to the following sufficient condition for the lossless transmission of a correlated source pair over a DMIC.
\begin{theorem}\label{th:1}
A source pair $(S,T)\sim p(s,t)$ is admissible for a discrete
memoryless interference channel $p(y_1y_2|x_1x_2)$ if there exist
auxiliary random variables $W_0,W_1,W_2$ with joint
distribution of all the variables factoring as:
\beq p(s,t,w_0,w_1,w_2,x_1,x_2)=p(s,t)p(w_0)p(w_1|w_0)p(w_2|w_0)p(x_1|sw_1w_0)p(x_2|tw_2w_0),\label{eq:dist2}\eeq
and the following conditions are satisfied:
\bqa H(S|K)&<&I(SX_1;Y_1|W_0W_2K),\\
H(T|K)&<&I(TX_2;Y_2|W_0W_1K),\\
H(S)&<&I(W_0W_2SX_1;Y_1),\\
H(T)&<&I(W_0W_1TX_2;Y_2),\\
H(S|K)+H(T|K)&<&I(SX_1;Y_1|W_0W_1W_2K)+I(W_1TX_2;Y_2|W_0K),\\
H(S|K)+H(T|K)&<&I(TX_2;Y_2|W_0W_1W_2K)+I(W_2SX_1;Y_1|W_0K),\\
H(S|K)+H(T|K)&<&I(SW_2X_1;Y_1|W_0W_1K)+I(TW_1X_2;Y_2|W_0W_2K),\\
H(S|K)+H(T)&<&I(W_0W_1TX_2;Y_2)+I(SX_1;Y_1|W_0W_1W_2K),\\
H(S)+H(T|K)&<&I(W_0W_2SX_1;Y_1)+I(TX_2;Y_2|W_0W_1W_2K),\\
2H(S|K)+H(T|K)&<&I(SX_1;Y_1|W_0W_1W_2K)+I(SW_2X_1;Y_1|W_0K)\nn\\
&&+I(TW_1X_2;Y_2|W_0W_2K),\\
H(S|K)+2H(T|K)&<&I(TX_2;Y_2|W_0W_1W_2K)+I(TW_1X_2;Y_2|W_0K)\nn\\
&&+I(SW_2X_1;Y_1|W_0W_1K),\\
H(S)+H(S|K)+H(T|K)&<&I(SX_1;Y_1|W_0W_1W_2K)+I(W_0W_2SX_1;Y_1)\nn\\
&&+I(TW_1X_2;Y_2|W_0W_2K),\\
H(T)+H(S|K)+H(T|K)&<&I(TX_2;Y_2|W_0W_1W_2K)+I(W_0W_1TX_2;Y_2)\nn\\
&&+I(SW_2X_1;Y_1|W_0W_1K),\eqa

where $K=f(S)=g(T)$ is the common part of $S$ and $T$ in the sense of G\'{a}cs, K\"{o}rner \cite{Gacs&Korner:1973} and Witsenhausen\cite{Witsenhausen:1975}.
\end{theorem}

We now discuss some implications of Theorem 1. 
\begin{corollary}
 If there is no common part for the source pair $(S,T)$, i.e., $K=\emptyset$,  let $W_0=Q$ be the time sharing variable. Theorem 1 yields the following sufficient condition for the lossless transmission of $(S,T)$ over a discrete memoryless interference channel.

 \bqa H(S)&<&I(SX_1;Y_1|QW_2),\\
H(T)&<&I(TX_2;Y_2|QW_1),\\
H(S)+H(T)&<&I(SX_1;Y_1|QW_1W_2)+I(W_1TX_2;Y_2|Q),\\
H(S)+H(T)&<&I(TX_2;Y_2|QW_1W_2)+I(W_2SX_1;Y_1|Q),\\
H(S)+H(T)&<&I(SW_2X_1;Y_1|QW_1)+I(TW_1X_2;Y_2|QW_2),\\
2H(S)+H(T)&<&I(SX_1;Y_1|QW_1W_2)+I(SW_2X_1;Y_1|Q)\nn\\
&&+I(TW_1X_2;Y_2|QW_2),\\
H(S)+2H(T)&<&I(TX_2;Y_2|QW_1W_2)+I(TW_1X_2;Y_2|Q)\nn\\
&&+I(SW_2X_1;Y_1|QW_1),\eqa

where $W_1,W_2$ are auxiliary random variables such that the joint distribution of all variables can be factored as
\beq p(s,t,q,w_1,w_2,x_1,x_2)=p(s,t)p(q)p(w_1|q)p(w_2|q)p(x_1|sw_1q)p(x_2|tw_2q).\eeq
\end{corollary}
The fact that Theorem 1 includes the HK region as its special case comes directly from Corollary 1. If $S$ and $T$ are independent, choose the joint distribution as
\beq p(s,t,q,w_1,w_2,x_1,x_2)=p(s)p(t)p(q)p(w_1|q)p(w_2|q)p(x_1|w_1q)p(x_2|w_2q),\eeq
and let $R_1=H(S)$ and $R_2=H(T)$. Corollary 1 yields an achievable region for the interference channel which coincides with that described in Proposition 3.

Consider now another special case where the source has a special correlation structure similar to that of \cite{Slepian&Wolf:73IT2}.

\begin{corollary} Suppose that the source $(S,T)$ can be decomposed into three parts: $S=(S',K)$ and $T=(T',K)$ where $S',T',K$ are independent random variables. Choose the joint distribution
\beq p(s,t,w_0,w_1,w_2,x_1,x_2)=p(s')p(t')p(k)p(w_0)p(w_1|w_0)p(w_2|w_0)p(x_1|w_0w_1)p(x_2|w_0w_2),\eeq 
where $s=(s',k)$ and $t=(t',k)$.  Theorem 1 gives the following sufficient condition for the lossless transmission of the source pair $(S,T)$.

\bqa H(S')&<&I(X_1;Y_1|W_0W_2),\\
H(T')&\leq&I(X_2;Y_2|W_0W_1),\\
H(K)+H(S')&<&I(W_0W_2X_1;Y_1),\\
H(K)+H(T')&<&I(W_0W_1X_2;Y_2),\\
H(S')+H(T')&<&I(X_1;Y_1|W_0W_1W_2)+I(W_1X_2;Y_2|W_0),\\
H(S')+H(T')&<&I(X_2;Y_2|W_0W_1W_2)+I(W_2X_1;Y_1|W_0),\\
H(S')+H(T')&<&I(W_2X_1;Y_1|W_0W_1)+I(W_1X_2;Y_2|W_0W_2),\\
H(K)+H(S')+H(T')&<&I(W_0W_1X_2;Y_2)+I(X_1;Y_1|W_0W_1W_2),\\
H(K)+H(S')+H(T')&<&I(W_0W_2X_1;Y_1)+I(X_2;Y_2|W_0W_1W_2),\\
2H(S')+H(T')&<&I(X_1;Y_1|W_0W_1W_2)+I(W_2X_1;Y_1|W_0)\nn\\
&&+I(W_1X_2;Y_2|W_0W_2),\\
H(S')+2H(T')&<&I(X_2;Y_2|W_0W_1W_2)+I(W_1X_2;Y_2|W_0)\nn\\
&&+I(W_2X_1;Y_1|W_0W_1),\\
H(K)+2H(S')+H(T')&<&I(X_1;Y_1|W_0W_1W_2)+I(W_0W_2X_1;Y_1)\nn\\
&&+I(W_1X_2;Y_2|W_0W_2),\\
H(K)+H(S')+2H(T')&<&I(X_2;Y_2|W_0W_1W_2)+I(W_0W_1X_2;Y_2)\nn\\
&&+I(W_2X_1;Y_1|W_0W_1).\eqa
\end{corollary}
Corollary 2 can be used to establish that the sufficient condition includes that of \cite{Jiang-etal:06Arsilomar,Cao&Chen&Zhang:WCNC07,Jiang&Xin:2008} as its special case. Specifically, define $R_0=H(K), R_1=H(S')$ and $R_2=H(T')$, the sufficient condition reduces to the rate region of interference channels with common information obtained in \cite{Jiang-etal:06Arsilomar,Cao&Chen&Zhang:WCNC07,Jiang&Xin:2008}.

\section{Separation is Strictly Suboptimal}
In this section, we give a simple example to show that separate source and channel coding is in general not optimal for sending correlated sources over interference channels.

Consider the transmission of a correlated binary source $(S, T)$ with the joint distribution $p(s,t)$ given by
\bqa p(s=0,t=0)&=&\frac{1}{3},\\
    p(s=0,t=1)&=&0,\\
    p(s=1,t=0)&=&\frac{1}{3},\\
    p(s=1,t=1)&=&\frac{1}{3},\eqa

over an interference channel defined by
\bqa {\cal X}_1&=&{\cal X}_2=\{0,1\},\\
{\cal Y}_1&=&\{0,1\},\\
{\cal Y}_2&=&\{0,1,2\},\\
Y_1&=&X_1,\\
Y_2&=&X_1+X_2.\eqa

The source is the same ``triangular'' source used in \cite{Cover_etal:1980} and \cite{Han&Costa:1987} to demonstrate that separation is not optimal for communicating correlated sources over MAC and BC, respectively. The channel is a special case of the deterministic interference channel studied in \cite[Theorem 2]{ElGamal&Costa:82IT} whose capacity region is the convex closure of those $(R_1,R_2)$ pairs satisfying
\bqa R_1&\leq&H(X_1),\\
     R_2&\leq&H(X_2),\\
     R_1+R_2&\leq&H(Y_2),\eqa
     over all the product probability distribution $p(x_1)p(x_2)$.

\begin{figure}[htp]
\centerline{\leavevmode \epsfxsize=3.05in \epsfysize=2.55in
\epsfbox{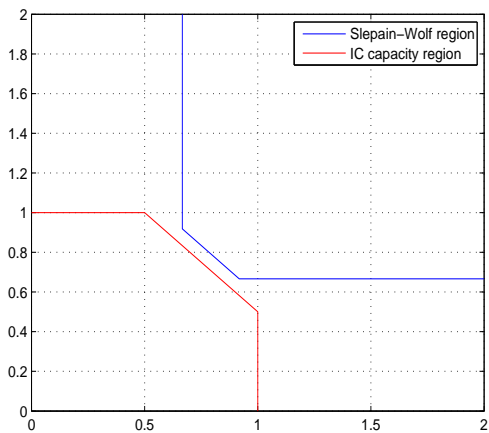}}
\caption{\label{fig:sw1} The Slepian-Wolf region and the IC capacity region.}
\end{figure}

It can be easily calculated that $H(S,T)=\log{3}=1.58$ bits. On the other hand, if $X_1$ and $X_2$ are independent, we have,
\beq R_1+R_2\leq \max_{p(x_1)p(x_2)}H(Y_2)=1.5 .\eeq
    Thus, $H(S,T)> H(Y_2)$ for all $p(x_1)p(x_2)$. Therefore, lossless transmission is not attainable by a simple concatenation of the Slepian-Wolf code followed by an optimal channel code. This is illustrated in Fig. \ref{fig:sw1} where the capacity region for the interference channel and the Slepian-Wolf rate region of the source pair do not intersect. However, it can be easily checked that a trivial way to reliably transmit this source is to choose $X_1=S$ and $X_2=T$, which results in zero error probability at both receivers. This example shows that separate source and channel coding is strictly suboptimal.

One can also easily check that this special case is included in Theorem \ref{th:1}. Let $K=W_0=W_1=W_2=\phi$, $X_1=S$, and $X_2=T$, the condition in Theorem \ref{th:1} reduces to
    \bqa H(S)&\leq& I(S;Y_1)\label{eq:ex1},\\
         H(T)&\leq& I(T;Y_2)\label{eq:ex2}.\eqa
  The first inequality is obviously satisfied  given that $Y_1=S$. To check the second inequality , note that since $Y_2=S+T$,
\beq
I(T;Y_2)=H(S+T)-H(S+T|T)=H(S+T)-H(S|T).
\eeq
Thus the second inequality reduces to 
    \beq H(S,T)\leq H(S+T)=\log{3},\eeq
    which is also satisfied for the given source pair.

\section{Conclusion}
  In this paper, we studied the problem of communicating an arbitrarily correlated source over a discrete memoryless interference channel. Using the techniques of correlation preserving coding and random source partition,  a sufficient condition was derived for reliable transmission of correlated sources over interference channels. The proposed region includes the Han and Kobayashi achievable rate region for general interference channels as its special case. Furthermore, it includes the known rate region for interference channels with common information as its special case when the course correlation is in the sense of \cite{Slepian&Wolf:73IT2}.  Finally, a simple example is given to show that separate source and channel coding is strictly suboptimal for communicating correlated sources over interference channels.

\section*{Appendix - Proof of Theorem 1 }
Theorem 1 can be obtained via Fourier-Mortzkin elimination from the following constraints.

\bqa
H(S|K)-r_1&<& I(SX_1;Y_1|W_0W_1W_2K)\label{eq:dec1},\\
H(S|K)&<&I(SX_1;Y_1|W_0W_2K),\\
H(S|K)-r_1+r_2&<&I(SW_2X_1;Y_1|W_0W_1K),\\
H(S|K)+r_2&<&I(W_2SX_1;Y_1|W_0K),\\
H(S)+r_2&<&I(W_0W_2SX_1;Y_1)\label{eq:dec2},\\
H(T|K)-r_2&<& I(TX_2;Y_2|W_0W_1W_2K)\label{eq:dec3},\\
H(T|K)&<&I(TX_2;Y_2|W_0W_1K),\\
H(T|K)-r_2+r_1&<&I(TW_1X_2;Y_2|W_0W_2K),\\
H(T|K)+r_1&<&I(W_1TX_2;Y_2|W_0K),\\
H(T)+r_1&<&I(W_0W_1TX_2;Y_2)\label{eq:dec4},\\
r_1&\geq&0,\\
r_2&\geq&0\label{eq:dec5}.\eqa

Therefore, it suffices to prove, for decoder 1, that equations (\ref{eq:dec1}-\ref{eq:dec2}) constitute a sufficient condition.  As sketched in Section III, the coding scheme involves Cover-El Gamal-Salehi's
\cite{Cover_etal:1980} correlation preserving coding and also Han and
Costa's \cite{Han&Costa:1987} random source partition.

a) {\em Random partition of the source sequences:}
 Let $r_1\geq 0, r_2\geq 0$ be any nonnegative real numbers.
  Randomly place source sequences $S^n \in {\cal S}^n$  into $2^{nr_1}$ cells  and
 denote the cell index for a given $s^n$ by $\alpha=l_1(s^n)\in I_1=\{1,2,\cdots, 2^{nr_1}\}$.
 Similarly, randomly place each $T^n\in {\cal T}^n$ into $2^{nr_2}$ cells and denote the cell index for a given $t^n$ by $\beta=l_2(t^n)\in I_2=\{1,2,\cdots,
 2^{nr_2}\}$. For this random source partition, we have the following lemma as in \cite{Han&Costa:1987}.
 \begin{lemma}
 Let ${\cal S}_0, {\cal T}_0$ be any subset of ${\cal S}^n$ and ${\cal T}^n$, respectively. Then for any $\alpha \in I_1$ and $\beta \in I_2$, we have,
 \bqa
 {\cal E}(|\{s^n\in {\cal S}_0\}: l_1(s^n)=\alpha|)&=&|{\cal S}_0|\times 2^{-nr_1},\\
 {\cal E}(|\{t^n\in {\cal T}_0\}: l_2(t^n)=\beta|)&=&|{\cal T}_0|\times 2^{-nr_2},\eqa
 where ${\cal E}\{\cdot\}$ denotes the expectation.
 \end{lemma}

b) {\em Codebook generation:} For any given joint distribution defined
 in (\ref{eq:dist2}), we first calculate the following
 distributions:
 $p(w_0),p(w_1|w_0)$, $p(w_2|w_0)$, $p(x_1|w_0w_1s)$ and $p(x_2|w_0w_2t)$.

 For each $k^n \in {\cal K}^n$, independently generate one $w^n_0$ sequence according to $\prod^n_{i=1}p(w_{0i})$. Index them by $w^n_0(k^n)$.
 For each source sequence $s^n$, find its cell index $\alpha$ and the corresponding auxiliary sequence $w^n_0(f(s^n))$, and independently generate {\em one} codeword $w^n_1$ according to $\prod^n_{i=1}p(w_{1i}|w_{0i})$. Index them by $w^n_1(\alpha, w^n_0)$. Next, for each $s^n$, find the corresponding $w^n_0(f(s^n))$ and $w^n_1(\alpha, w^n_0)$, independently generate one codeword $x^n_1$ according to $\prod^n_{i=1}p(x_{1i}|w_{0i}w_{1i}s_{i})$. Index them by $x^n_1(s^n,w^n_0,w^n_1)$. Similarly, generate codewords $w^n_2(\beta, w^n_0)$ and $x^n_2(t^n,w^n_0,w^n_2)$ for user 2.

 Notice that, for user $1$, there are three set of codewords: $w^n_0(k^n), w^n_1(\alpha,w^n_0)$ and $x^n_1(s^n,w^n_0,w^n_1)$. Here, $w^n_0(k^n)$ carries the information corresponding to the common part of the sources; $w^n_1(\alpha,w^n_0)$ carries the information of the cell index of the source $s^n$ which is superimposed on top of $w^n_0$;  $x^n_1(s^n,w^n_0,w^n_1)$ carries the private information of the source $s^n$ and is superimposed on top of both $w^n_0$ and $w^n_1$. The codebook structure for user 2 is similar.

c) {\em Encoding:} Upon observing $s^n$, encoder 1 finds the cell
index $\alpha=l_1(s^n)$, codewords $w^n_0(f(s^n))$ and $w^n_1(\alpha, w^n_0)$, and then sends
the corresponding $x^n_1(s^n,w^n_0,w^n_1)$. Similarly, encoder 2 sends
$x^n_2(t^n,w^n_0,w^n_2)$.

d) {\em Decoding:} Upon receiving $y^n_1$, decoder 1 declares
$\hat{s}^n=s^n$ to be the transmitted source sequence if $s^n$
is the unique sequence such that
\beq (s^n,k^n,w^n_0,w^n_1,x^n_1,w^n_2,y^n_1)\in
T^n_{\epsilon}(SKW_0W_1X_1W_2Y_1).\eeq
 Similarly, decoder 2 finds the
unique $t^n$ such that \beq(t^n,k^n,w^n_0,w^n_2,x^n_2,w^n_1,y^n_2)\in
T^n_{\epsilon}(TKW_0W_2X_2W_1Y_2).\eeq

e) {\em Error analysis:} Encoding error occurs only if the source sequence pairs $(s^n,t^n)\notin T^n_{\epsilon}(ST)$ whose probability is bounded by $\epsilon$ from the asymptotic equipartition property (AEP)\cite{Cover:book}. By symmetry, we only need to consider the decoding errors at receiver 1.

Suppose $(s^n_0, t^n_0)\in T^n_{\epsilon}$ are the source outputs, with $k^n_0=f(s^n_0)=g(t^n_0)$. Without loss of generality, we assume that $\alpha=1$ and
$\beta=1$, i.e., $l_1(s^n_0)=1$ and $ l_2(t^n_0)=1$, and also $w^n_0(k^n_0)=w^n_{00}$. Then
an error occurs if any one of the following events happens:
\begin{enumerate}

\item $E_{11}:  \Big(s^n_0,k^n_0,w^n_0(k^n_0),w^n_1(1,w^n_0), x^n_1(s^n_0,w^n_0,w^n_1),w^n_2(1,w^n_0),y^n_1\Big)\notin T^n_{\epsilon}(SKW_0W_1X_1W_2Y_1).$

 \item $E_{12}$: there exists some $s^n\neq s^n_0$ in the cell $\alpha=1$, i.e., $l_1(s^n)=1$, and $\beta=1$, such that,
  $\Big(s^n,k^n_0,w^n_0(k^n_0),w^n_1(1,w^n_0),x^n_1(s^n,w^n_0,w^n_1),w^n_2(1,w^n_0),y^n_1\Big)\in T^n_{\epsilon}(SKW_0W_1X_1W_2Y_1).$

\item $E_{13}$: there exists some $s^n\neq s^n_0$ in the cell $\alpha=1$, i.e., $l_1(s^n)=1$, and $\beta\neq 1$, such that,

  $\Big(s^n,k^n_0,w^n_0(k^n_0),w^n_1(1,w^n_0), x^n_1(s^n,w^n_0,w^n_1),w^n_2(\beta,w^n_0),y^n_1\Big)\in T^n_{\epsilon}(SKW_0W_1X_1W_2Y_1).$

\item $E_{14}$: there exists some $s^n\neq s^n_0$ in the cell $\alpha\neq 1$, i.e., $l_1(s^n)=\alpha$, and $\beta=1$, such that,

  $\Big(s^n,k^n_0,w^n_0(k^n_0),w^n_1(\alpha,w^n_0), x^n_1(s^n,w^n_0,w^n_1),w^n_2(1,w^n_0),y^n_1\Big)\in T^n_{\epsilon}(SKW_0W_1X_1W_2Y_1).$

\item $E_{15}$: there exists some $s^n\neq s^n_0$ in the cell $\alpha\neq 1$, i.e., $l_1(s^n)=\alpha$, and $\beta\neq1$, such that,

  $\Big(s^n,k^n_0,w^n_0(k^n_0),w^n_1(\alpha,w^n_0), x^n_1(s^n,w^n_0,w^n_1),w^n_2(\beta,w^n_0),y^n_1\Big)\in T^n_{\epsilon}(SKW_0W_1X_1W_2Y_1).$

\item $E_{16}$: there exists some $s^n\neq s^n_0$ in the cell $\alpha\neq 1$, i.e., $l_1(s^n)=\alpha$, and $\beta\neq1$, such that,

$k^n=f(s^n)\neq k^n_0$, $w^n_0(k^n)\neq w^n_0(k^n_0)$\\
 and $\Big(s^n,k^n,w^n_0(k^n),w^n_1(\alpha,w^n_0), x^n_1(s^n,w^n_0,w^n_1),w^n_2(\beta,w^n_0),y^n_1\Big)\in T^n_{\epsilon}(SKW_0W_1X_1W_2Y_1).$

\item $E_{17}$: there exists some $s^n\neq s^n_0$ in the cell $\alpha\neq 1$, i.e., $l_1(s^n)=\alpha$, and $\beta\neq1$, such that,

$k^n=f(s^n)\neq k^n_0$, $w^n_0(k^n)= w^n_0(k^n_0)$\\
and $\Big(s^n,k^n,w^n_0(k^n),w^n_1(\alpha,w^n_0), x^n_1(s^n,w^n_0,w^n_1),w^n_2(\beta,w^n_0),y^n_1\Big)\in T^n_{\epsilon}(SKW_0W_1X_1W_2Y_1).$
\end{enumerate}

Hence, the probability of error at decoder 1 is
\beq P_{e1}=Pr\{\cup^7_{i=1} E_{1i}\}\leq \sum^7_{i=1}Pr\{E_{1i}\}.\eeq

We evaluate the probabilities of the seven error events individually. First, by the AEP, 
\beq Pr\{E_{11}\}\leq \epsilon,\eeq for sufficiently large $n$.
For the second event, we have,\footnote{For notational ease, we use $T^n_{\epsilon}$ to denote the typical set $T^n_{\epsilon}(SKW_0W_1X_1W_2Y_1)$ below.}

\bqa &&Pr\{E_{12}\}\nn\\
&=&{\cal E}\Big(\sum_{s^n\in {\cal L}_1(1)\cap T^n_{\epsilon}(S)}Pr\Big((s^n,k^n_0,w^n_0(k^n_0),w^n_1(1,w^n_0),x^n_1(s^n,w^n_0,w^n_1),w^n_2(1,w^n_0),y^n_1)\in T^n_{\epsilon}\Big),\nn\\
                 &=&{\cal E}\Big(\sum_{s^n\in {\cal L}_1(1)\cap T^n_{\epsilon}(S)}\sum_{(s^n,k^n,w^n_0,w^n_1,x^n_1,w^n_2,y^n_1)\in T^n_{\epsilon}}p(k^nw^n_0w^n_1w^n_2y^n_1)p(s^n)p(x^n_1|s^nw^n_1w^n_0)\Big),\nn\\
                 &\leq&{\cal E}\Big(\sum_{s^n\in {\cal L}_1(1)\cap T^n_{\epsilon}(S)}|T^n_{\epsilon}(SKW_0W_1X_1W_2Y_1)|2^{-n(H(KW_0W_1W_2Y_1)}2^{n(H(S)+H(X_1|SW_1W_0)-3\epsilon)}\Big),\nn\\
                 &\leq&{\cal E}\Big(\sum_{s^n\in {\cal L}_1(1)\cap T^n_{\epsilon}(S)}2^{n(H(SKW_0W_1X_1W_2Y_1)-H(KW_0W_1W_2Y_1))}2^{-n(H(S)+H(X_1|SW_1W_0)-4\epsilon)}\Big),\nn\\
                 &=&{\cal E}\Big(\sum_{s^n\in {\cal L}_1(1)\cap T^n_{\epsilon}(S)}2^{n(H(S|KW_0W_1W_2)+H(X_1Y_1|SKW_0W_1W_2))},\nn\\
                 &&\cdot 2^{-n(H(Y_1|KW_0W_1W_2)+H(S)+H(X_1|SW_1W_0)-4\epsilon)}\Big),\nn\\
                 &=&{\cal E}\Big(\sum_{s^n\in {\cal L}_1(1)\cap T^n_{\epsilon}(S)}2^{n(H(S|K)+H(Y_1|SKW_0W_1W_2X_1))}
                 2^{-n(H(S)+H(Y_1|KW_0W_1W_2)-4\epsilon)}\Big),\nn\\
               &\leq&2^{n(H(S)+\epsilon-r_1)}2^{-n(I(SX_1;Y_1|KW_0W_1W_2)+I(S;K))+4\epsilon)},\nn\\
               &=&2^{n(H(S|K)-r_1-I(SX_1;Y_1|KW_0W_1W_2)+5\epsilon)}\label{eq:er12}.\eqa

For the third event, we have,
\bqa &&Pr\{E_{13}\}\nn\\
&=&{\cal E}\Big(\sum_{s^n\in {\cal L}_1(1)\cap T^n_{\epsilon}(S)}\sum_{\beta\neq1}Pr((s^n,k^n_0,w^n_0,w^n_1(1,w^n_0),x^n_1(s^n,w^n_1,w^n_0),w^n_2(\beta,w^n_0),y^n_1)\in T^n_{\epsilon})\Big),\nn\\
&=&{\cal E}\Big(\sum_{s^n\in {\cal L}_1(1)\cap T^n_{\epsilon}(S)}\sum_{\beta\neq1}\sum_{(s^n,k^n,w^n_0,w^n_1,x^n_1,w^n_2,y^n_1)\in T^n_{\epsilon}}p(s^n)p(w^n_2|w^n_0)p(x^n_1|s^nw^n_0w^n_1)p(k^nw^n_0w^n_1y^n_1)\Big),\nn\\
&\leq&{\cal E}\Big(\sum_{s^n\in {\cal L}_1(1)\cap T^n_{\epsilon}(S)}\sum_{\beta\neq1}2^{n(H(SKW_0W_1W_2X_1Y_1)}2^{-n(H(S)+H(W_2|W_0)+H(X_1|W_1W_0S)+H(KW_0W_1Y_1)-5\epsilon)},\nn\\
&=&{\cal E}\Big(\sum_{s^n\in {\cal L}_1(1)\cap T^n_{\epsilon}(S)}\sum_{\beta\neq1}2^{n(H(S|KW_0W_1)+H(W_2X_1Y_1|SKW_0W_1))},\nn\\
&&\cdot 2^{-n(H(S)+H(W_2|W_0)+H(X_1|W_1W_0S)+H(Y_1|KW_0W_1)-5\epsilon)}\Big),\nn\\
&\leq&2^{n(H(S)+\epsilon-r_1+r_2)}2^{-n(I(S;K)+I(SW_2X_1;Y_1|KW_0W_1)-5\epsilon)},\nn\\
&=&2^{n(H(S|K)-r_1+r_2-I(SW_2X_1;Y_1|KW_0W_1)+6\epsilon)}.\label{eq:er13}\eqa

For the fourth event, we have,

\bqa &&Pr\{E_{14}\}\nn\\
&=&{\cal E}\Big(\sum_{\alpha\neq1}\sum_{s^n\in {\cal L}_1(\alpha)\cap T^n_{\epsilon}(S)}Pr((s^n,k^n_0,w^n_0,w^n_1(\alpha,w^n_0),x^n_1(s^n,w^n_1,w^n_0),w^n_2(1,w^n_0),y^n_1)\in T^n_{\epsilon})\Big),\nn\\
&=&{\cal E}\Big(\sum_{\alpha\neq1}\sum_{s^n\in {\cal L}_1(\alpha)\cap T^n_{\epsilon}(S)}\sum_{(s^n,k^n,w^n_0,w^n_1,x^n_1,w^n_2,y^n_1)\in T^n_{\epsilon}}p(s^n)p(w^n_1|w^n_0)p(x^n_1|s^nw^n_0w^n_1)p(w^n_0)p(k^nw^n_2y^n_1|w^n_0)\Big),\nn\\
&\leq&2^{n(H(S)+\epsilon)}2^{n(H(SKW_0W_1W_2X_1Y_1)}2^{-n(H(SW_0W_1X_1)+H(KW_2Y_1|W_0)+6\epsilon)},\nn\\
&=&2^{n(H(S)-I(SW_1X_1;KW_2Y_1|W_0)+7\epsilon)},\nn\\
&=&2^{n(H(S)-I(SW_1X_1;W_2|W_0)-I(SX_1;Y_1K|W_0W_2))}2^{-n(I(W_1;Y_1|W_0W_2X_1S)-7\epsilon)},\nn\\
&\stackrel{(a)}{=}&2^{n(H(S)-I(SX_1;Y_1K|W_0W_2)+7\epsilon)},\nn\\
&=&2^{n(H(S)-I(S;K|W_0W_2)-I(X_1;K|W_0W_2S))}2^{-n(I(SX_1;Y_1|KW_0W_2)-7\epsilon)},\nn\\
&=&2^{n(H(S|K)-I(SX_1;Y_1|KW_0W_2)+7\epsilon)}\label{eq:er14},\eqa
where $(a)$ is because $W_2\rightarrow W_0 \rightarrow SW_1X_1$ and $W_1\rightarrow SW_2X_1\rightarrow Y_1$ form Markov chains.

For the fifth event, we have,

\bqa &&Pr\{E_{15}\}\nn\\
&=&{\cal E}\Big(\sum_{\alpha\neq1}\sum_{s^n\in {\cal L}_1(\alpha)\cap T^n_{\epsilon}(S)}\sum_{\beta\neq1}Pr((s^n,k^n_0,w^n_0,w^n_1(\alpha,w^n_0),x^n_1(s^n,w^n_1,w^n_0),w^n_2(\beta,w^n_0),y^n_1)\in T^n_{\epsilon})\Big),\nn\\
&=&{\cal E}\Big(\sum_{\alpha\neq1}\sum_{s^n\in {\cal L}_1(\alpha)\cap T^n_{\epsilon}(S)}\sum_{\beta\neq1}\sum_{(s^n,k^n,w^n_0,w^n_1,x^n_1,w^n_2,y^n_1)\in T^n_{\epsilon}}
p(s^n)p(w^n_1|w^n_0)p(w^n_2|w^n_0)p(x^n_1|s^nw^n_0w^n_1)p(k^nw^n_0y^n_1)\Big),\nn\\
&\leq&2^{n(H(S)+r_2+\epsilon)}2^{n(H(SKW_0W_1W_2X_1Y_1)-H(KW_0Y_1))}
2^{-n(H(S)+H(W_1W_2|W_0)+H(X_1|SW_0W_1)-6\epsilon)},\nn\\
&=&2^{n(H(S)+r_2+H(SKX_1Y_1|W_0W_1W_2))}
2^{-n(H(KY_1|W_0)+H(S)+H(X_1|SW_0W_1)-7\epsilon)},\nn\\
&=&2^{n(H(S)+r_2+H(SX_1Y_1|KW_0W_1W_2))}
2^{-n(H(Y_1|KW_0)+H(S)+H(X_1|SW_0W_1)-7\epsilon)},\nn\\
&=&2^{n(H(S)+r_2+H(S|K)+H(X_1Y_1|SKW_0W_1W_2))}
2^{-n(H(S)+H(Y_1|KW_0)+H(X_1|SW_0W_1)+7\epsilon)},\nn\\
&=&2^{n(H(S|K)+r_2-I(SW_1W_2X_1;Y_1|KW_0)+7\epsilon)},\nn\\
&=&2^{n(H(S|K)+r_2-I(W_2SX_1;Y_1|KW_0))}2^{-n(I(W_1;Y_1|SKW_0W_2X_1)-7\epsilon)},\nn\\
&=&2^{n(H(S|K)+r_2-I(W_2SX_1;Y_1|KW_0)+7\epsilon)}.\label{eq:er15}\eqa

For the sixth event, we have,
\bqa
&&Pr\{E_{16}\}\nn\\
&=&{\cal E}\Big(\sum_{\alpha\neq1}\sum_{s^n\in {\cal L}_1(\alpha)\cap T^n_{\epsilon}(S)}\sum_{\beta\neq1}, \nn\\ &&
Pr(s^n,k^n,w^n_0(k^n),w^n_1(\alpha,w^n_0),x^n_1(s^n,w^n_1,w^n_0),w^n_2(\beta,w^n_0),y^n_1)\in T^n_{\epsilon}); w^n_0(k^n)\neq w^n_{00})\Big),\nn\\
&=&{\cal E}\Big(\sum_{\alpha\neq1}\sum_{s^n\in {\cal L}_1(\alpha)\cap T^n_{\epsilon}(S)}\sum_{\beta\neq1}\sum_{w^n_0\in{\cal W}^n_0}p(w^n_0={w'^n_0})\nn\\
&&\cdot Pr(s^n,k^n,w'^n_0(k^n),w^n_1(\alpha,w'^n_0),x^n_1(s^n,w^n_1,w'^n_0),
w^n_2(\beta,w'^n_0),y^n_1)\in T^n_{\epsilon})|w'^n_0\neq w^n_{00})\Big),\nn\\
&=&{\cal E}\Big(\sum_{\alpha\neq1}\sum_{s^n\in {\cal L}_1(\alpha)\cap T^n_{\epsilon}(S)}\sum_{\beta\neq1}\sum_{(s^n,k^n,w^n_0,w^n_1,x^n_1,w^n_2,y^n_1)\in T^n_{\epsilon}}
\sum_{w^n_0\in T^n_{\epsilon}(W_0)}p(w'^n_0)\nn\\
&&p(s^nk^nw'^n_0w^n_1w^n_2x^n_1)p(y^n_1)\Big),\nn\\
&\leq&2^{n(H(S)+r_2+\epsilon)}2^{n(H(W_0)+\epsilon)}2^{-n(H(W_0)+\epsilon)}2^{n(H(SKW_0W_1W_2X_1Y_1)-H(SKW_0W_1W_2X_1)-H(Y_1)+3\epsilon)}\nn\\
&=&2^{n(H(S)+r_2-I(SKW_0W_1W_2X_1;Y_1)+4\epsilon)},\nn\\
&=&2^{n(H(S)+r_2-I(SW_0W_2X_1;Y_1)-I(W_1;Y_1|SW_0W_2X_1)+4\epsilon)},\nn\\
&=&2^{n(H(S)+r_2-I(SW_0W_2X_1;Y_1)+4\epsilon)}.\label{eq:er16}\eqa

For the last event, we have,

 \bqa
&&Pr\{E_{17}\}\nn\\
&=&{\cal E}\Big(\sum_{\alpha\neq1}\sum_{s^n\in {\cal L}_1(\alpha)\cap T^n_{\epsilon}(S)}\sum_{\beta\neq1}Pr(s^n,k^n,w^n_0(k^n),w^n_1(\alpha,w^n_0),x^n_1(s^n,w^n_1,w^n_0),\nn\\
&&w^n_2(\beta,w^n_0),y^n_1)\in T^n_{\epsilon}; w^n_0(k^n)=w^n_{00})\Big),\nn\\
&=&{\cal E}\Big(\sum_{\alpha\neq1}\sum_{s^n\in {\cal L}_1(\alpha)\cap T^n_{\epsilon}(S)}\sum_{\beta\neq1}\sum_{w'^n_0\in{\cal W}^n_0}p(w'^n_0)p(w'^n_0=w^n_{00})
Pr(s^n,k^n,w'^n_0(k^n),w^n_1(\alpha,w'^n_0),\nn\\&&x^n_1(s^n,w^n_1,w'^n_0),
w^n_2(\beta,w'^n_0),y^n_1)\in T^n_{\epsilon}| w'^n_0=w^n_{00})\Big),\nn\\
&=&{\cal E}\Big(\sum_{\alpha\neq1}\sum_{s^n\in {\cal L}_1(\alpha)\cap T^n_{\epsilon}(S)}\sum_{\beta\neq1}\sum_{w'^n_0\in T^{\epsilon}_n(W_0)}p(w'^n_0)p(w'^n_0=w^n_{00})
Pr(s^n,k^n,w'^n_0(k^n),w^n_1(\alpha,w'^n_0),\nn\\&&x^n_1(s^n,w^n_1,w'^n_0),
w^n_2(\beta,w'^n_0),y^n_1)\in T^n_{\epsilon}| w'^n_0=w^n_{00})\Big),\nn\\
&=&{\cal E}\Big(\sum_{\alpha\neq1}\sum_{s^n\in {\cal L}_1(\alpha)\cap T^n_{\epsilon}(S)}\sum_{\beta\neq1}\sum_{(s^n,k^n,w^n_0,w^n_1,x^n_1,w^n_2,y^n_1)\in T^n_{\epsilon}}
\sum_{w'^n_0\in T^n_{\epsilon}(W_0)}p(w'^n_0)p(w'^n_0=w^n_{00})\nn\\
&&p(s^nk^nw^n_1w^n_2x^n_1|w^n_0)p(w^n_0y^n_1)\Big),\nn\\
&\leq&2^{n(H(S)+r_2+\epsilon)}2^{n(H(W_0)+\epsilon)}2^{-n(H(W_0-\epsilon))}2^{-n(H(W_0)-\epsilon)}
2^{n(H(SKW_0W_1W_2X_1Y_1)-H(SKW_0W_1W_2X_1)-H(Y_1|W_0)+3\epsilon)},\nn\\
&=&2^{n(H(S)+r_2-H(W_0)-I(SKW_1W_2X_1;Y_1|W_0)+7\epsilon)},\nn\\
&=&2^{n(H(S)+r_2-H(W_0)-I(SW_2X_1;Y_1|W_0)+7\epsilon)}.\label{eq:er17}\eqa

From (\ref{eq:er12})-(\ref{eq:er17}), if the following conditions are satisfied, then the probability of error at decoder 1 will vanish as $n$ goes to infinity.

\bqa H(S|K)-r_1&<& I(SX_1;Y_1|W_0W_1W_2K)\label{eq:fm1},\\
H(S|K)&<&I(SX_1;Y_1|W_0W_2K),\\
H(S|K)-r_1+r_2&<&I(SW_2X_1;Y_1|W_0W_1K),\\
H(S|K)+r_2&<&I(W_2SX_1;Y_1|W_0K),\\
H(S)+r_2&<&I(W_0W_2SX_1;Y_1),\label{eq:sum}\\
H(S)+r_2&<&I(W_2SX_1;Y_1|W_0)+H(W_0).\label{eq:redu}\eqa

One can easily check that (\ref{eq:redu}) is dominated by (\ref{eq:sum}), since,
\bqa I(W_2SX_1;Y_1|W_0)+H(W_0)&=&H(Y_1|W_0)+H(W_0)-H(Y_1|SW_0W_2X_1),\\
&\geq&H(Y_1)-H(Y_1|SW_0W_2X_1),\\
&=&I(W_0W_2SX_1;Y_1).\eqa

This establishes (\ref{eq:dec1}-\ref{eq:dec2}). Similarly, (\ref{eq:dec3}-\ref{eq:dec4}) can be established symmetrically for decoder 2. 
The proof of Theorem 1 is complete by applying the Fourier-Mortzkin elimination to (\ref{eq:dec1}-\ref{eq:dec5}).

\end{document}